# Metastable Monolayer Formation through a Connector Structure

AUTHOR NAMES

Simon B. Hollweger[1]

Anna Werkovits[1]

Oliver T. Hofmann[1]

AUTHOR ADDRESS

[1]Institute of Solid State Physics, Graz University of Technology, Graz, Austria

AUTHOR INFORMATION

**Corresponding Author**

Oliver T. Hofmann, Email: o.hofmann@tugraz.at

ABSTRACT

The intentional growth of metastable surface structures of organic molecules adsorbed on inorganic substrates is a challenging task. It is usually unclear which kinetic mechanism leads to the metastable surface polymorph after a deposition experiment. In this work we investigate a growth procedure that allows to intentionally grow a defined metastable surface structure starting from thermodynamic equilibrium. This procedure is applicable to organic-inorganic interface systems that exhibit a thermodynamically stable *connector* structure that can be exploited to grow the metastable target structure. With specific temperature and pressure changes in the system a significant yield of the target polymorph can be achieved. We demonstrate this procedure on a simplified microscopic interface system of rectangular molecules adsorbing on a square lattice substrate with kinetic Monte Carlo growth simulations.

**TOC GRAPHICS**

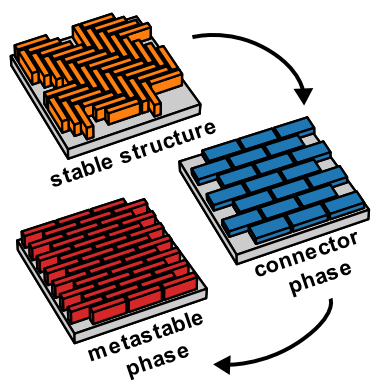

TEXT

It is well-known that the structure of monolayers of organic molecules adsorbed on metal substrates show pronounced polymorphism depending on their deposition conditions.[1–6] Controlling which structure the molecules on the surface form by varying certain deposition parameters like temperature and pressure is of great interest, as many interface properties, such as the injection barrier[7], charge-carrier mobility[8], work function and interface dipole,[9] depend significantly on the spatial arrangement of the first monolayer of the adsorbates.[10] However, targeted growth of a certain surface configuration of organic molecules is a highly challenging task, particularly when the desired structure is a metastable polymorph and thus thermodynamically unfavorable.

In this work, we suggest a way to overcome this difficulty by proposing a mechanism for growing a specifically targeted metastable surface structure starting from thermodynamic equilibrium. This procedure is based on two physical/chemical principles: First, in thermodynamic equilibrium the structure that emerges on the surface is the one that minimizes the Gibbs free energy of adsorption per area $\gamma$, which is commonly approximated[11] by

$$\gamma_k = \frac{1}{A_k}\left(E_k^{\text{ads}} - \mu(T,p)N_k\right) \tag{1}$$

where $E_k^{\text{ads}}$ is the adsorption energy of structure $k$, $\mu(T,p)$ the chemical potential of the molecular gas reservoir in contact with the surface at temperature $T$ and partial pressure $p$, and $N_k$ the number of adsorbed molecules in the unit cell with area $A_k$ (details can be found in the Supporting Information). Due to the linear dependence of $\gamma$ on $\mu$, it is possible to change the thermodynamically most stable surface structure by a targeted change of the chemical potential of the molecular gas reservoir by the variation of the temperature and pressure in the system. The second principle is Ostwald's rule of stages.[12] It states that during crystal formation, phases that are geometrically more similar to the initial phase are formed first, before transforming into the thermodynamically stable structure.

With these two concepts in mind, the conditions that allow systematically growing a metastable surface structure starting from thermodynamic equilibrium can be identified. One key feature of these systems is that they need to show the specific Gibbs free energy curve constellation depicted in Figure 1. It consists of three different surface structures. One of them is the metastable target structure (T) and the other two (S and C) are respective thermodynamically stable polymorphs at certain intervals of the chemical potential $\mu$. The decisive feature of this system is the existence of the connector structure (C). This structure must exhibit a different slope in the Gibbs free energy curve compared to the others and, therefore, intersects them at certain chemical potentials $\mu$. Equation (1) shows that the coverage $N_k/A_k$ of the structure determines the slope of the Gibbs curve $\gamma_k$, i.e., more densely packed structures exhibit steeper Gibbs curves. To conceptually realize this Gibbs curve constellation the surface structures S and T need to be relatively dense, and the connector phase C more loosely packed. To grow the metastable structure, two steps are now required, as indicated in Figure 1:

1. The first step is to create the thermodynamic stable connector phase C on the surface. This can be achieved by setting the chemical potential of the molecular reservoir to the region where the connector structure C is stable. Since this polymorph is a loosely packed structure it is stable at elevated temperatures. This facilitates the transition into the thermodynamic equilibrium on the surface experimentally, for example by using techniques such as hot-wall epitaxy.[13] It is also possible to transform the system into structure C with a post-deposition thermal treatment[6] where first another thermodynamic stable polymorph S is formed and with a temperature increase and a pressure drop a phase transition to the connector structure C is triggered. We will demonstrate this latter technique in a kinetic Monte Carlo growth simulation later in this work.

2. As soon as the connector structure C has formed, $\mu$ is increased to a region where polymorph C becomes unstable. Here, Ostwald's rule of stages comes into play. The system is likely to prefer a transition to the metastable target polymorph T rather than forming the thermodynamic stable polymorph S, if the connector phase C is geometrically more similar to the target polymorph T than the stable polymorph S.

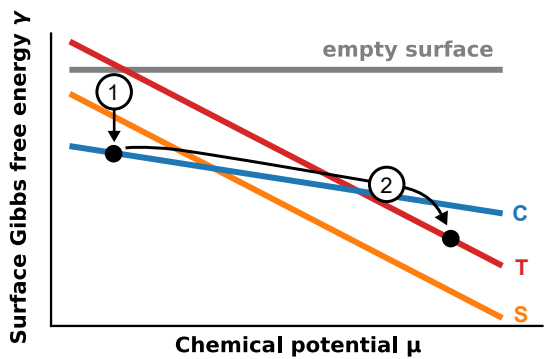

**Figure 1.** Schematic Gibbs free energy of adsorption per area over gas reservoir chemical potential. The two steps of growing the metastable structure T: (1) The first step is the formation of the connector phase C, (2) the second step is changing the chemical potential to trigger a transformation to the metastable target polymorph T.

As a proof of principle, we apply the described procedure of forming a metastable molecular monolayer to a hypothetical microscopic interface model in the framework of kinetic Monte Carlo (kMC).[14,15] The model consists of rectangular planar molecules adsorbing on a two-dimensional square lattice. The molecules can adsorb in four distinct adsorption geometries, two lying (face-on) and two standing (long edge-on) orientations, as shown in Figure 2a. The energy barriers between these adsorption geometries are shown in Figure 2c. The chemical potential $\mu$ of the molecular gas reservoir is approximated by an ideal gas.[11] To connect $\mu$ to pressures and temperatures, typical physical properties of organic molecules were used. Exemplarily, we took the mass, symmetry, and moments of inertia of 9-10 anthraquinone (9,10-dioxoanthracene), but we note that the qualitative findings remain the same for all small organic molecules. The details of the calculation are presented in the Supporting Information.

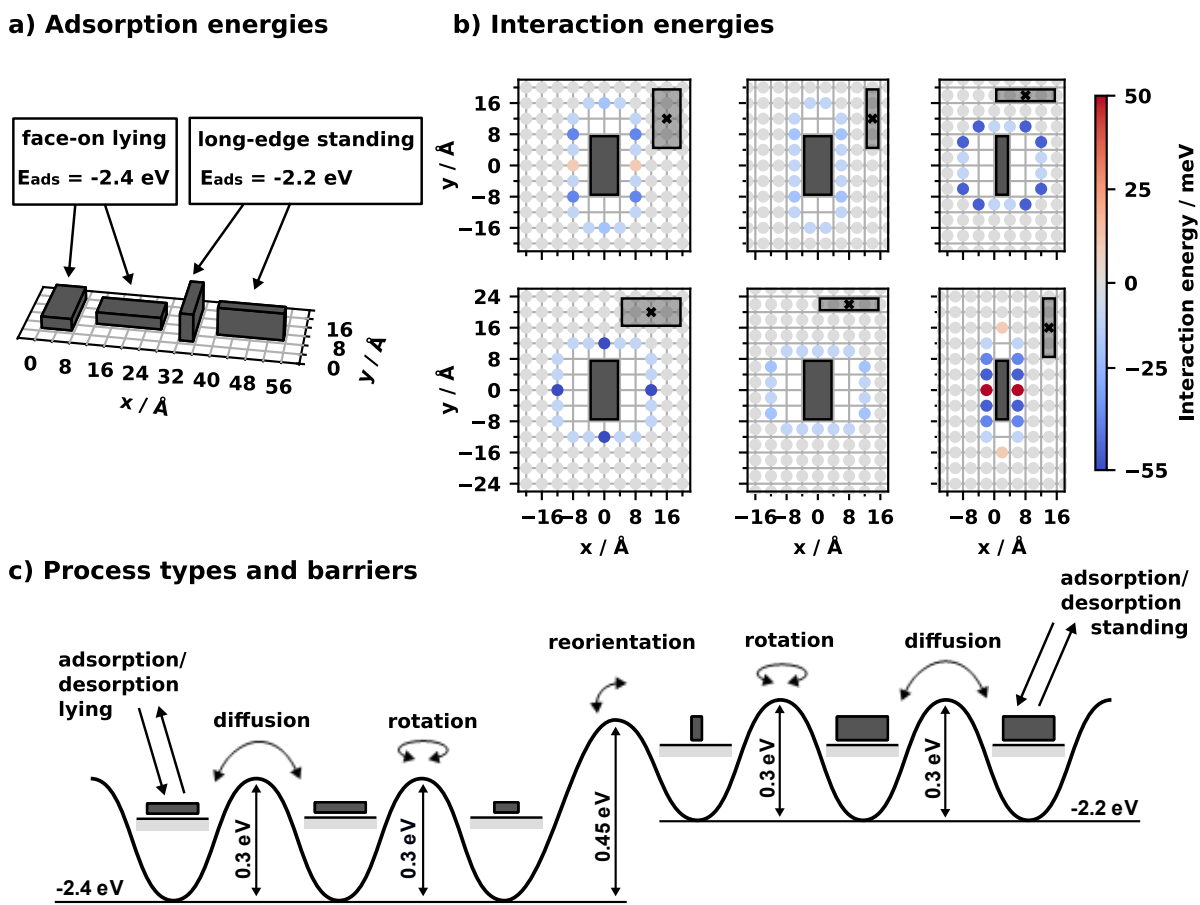

**Figure 2.** a) Adsorption geometries and energies of the rectangular molecules on a square grid with a lattice constant of 4 Å. b) Pair interaction energies of all combinations of the four adsorption geometries. Each point is a possible position for the geometric center of a neighbor molecule. The color of the points represents the interaction energy between the molecule pair. c) Barriers of the elementary processes (adsorption, desorption, diffusion, rotation and reorientation) between different adsorption geometries. The attempt frequency (pre-factor of the Arrhenius equation) of each on-surface process is set to $10^{12}$ $s^{-1}$.

The interactions between two long-edge standing molecules are based on typical interactions of π-conjugated systems.[16] These interactions penalize a direct overlap of the π-orbitals of the aromatic rings (which would result in Pauli-repulsion) and, therefore, prefer a shifted parallel or a perpendicular orientation between two molecules (Figure 2b). In this hypothetical model, there are two standing surface structures, a standing brickwall (SBW*) and a standing herringbone (SHB) pattern for the first monolayer, as shown in Figure 3.

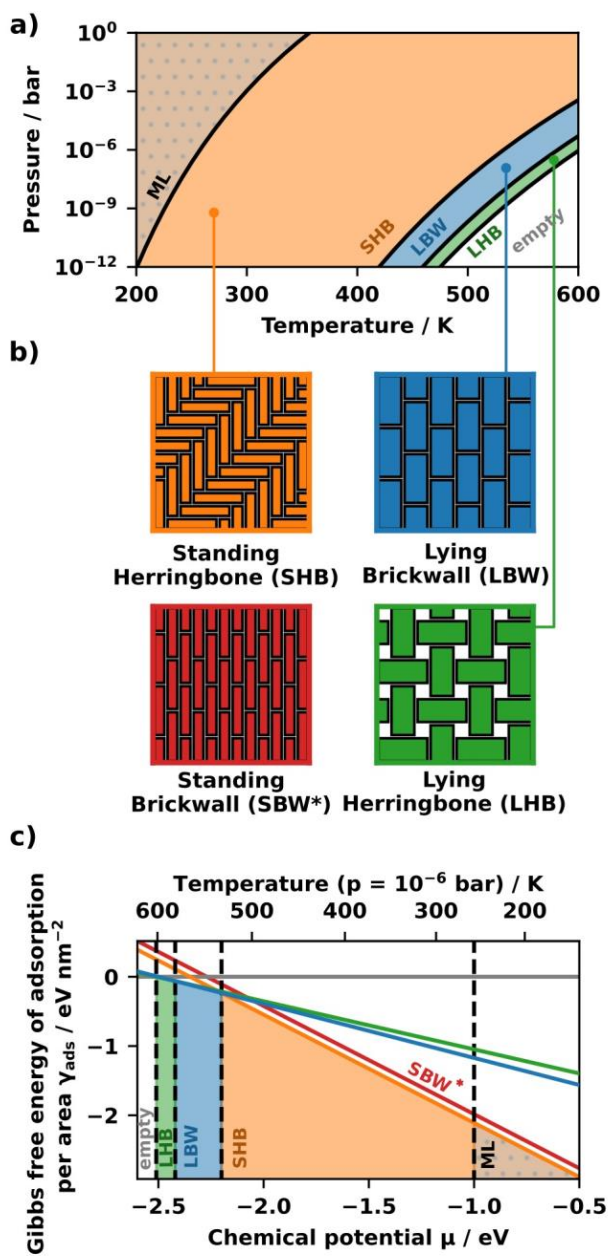

**Figure 3.** a) Thermodynamic phase diagram of the monolayer structures of the model system based on Equation (1). The same color coding as in Figure 1 is used. The dotted area is the multilayer region (ML) b) The four surface structures of the model. c) The Gibbs free energy of adsorption per area plotted over the chemical potential of the molecular gas reservoir. The filled area corresponds to the thermodynamically stable structure at this chemical potential range.

The lying molecules can also arrange in a brickwall (LBW) and a herringbone (LHB) pattern.

Additionally, we estimate the multilayer (ML) regime of the system by assuming that a molecule adsorbing in the second layer assumes an adsorption energy of $-1$ eV, which is a typical value for small organic molecules.[17] This leads to the dotted area in Figure 3.

Three of the monolayer structures (LBW, LHB, and SHB) are thermodynamically stable, as shown in the $p$-$T$ and $\gamma_{ads}$ vs. $\mu$ phase diagram in Figure 3. At room temperature, the phase diagram shows that the standing herringbone structure (SHB) is thermodynamically stable. The two lying phases, brick-wall (LBW) and herringbone (LHB), are stable at higher temperatures. The standing brickwall structure (SBW*) is only kinetically accessible, making it a metastable phase (indicated by the asterisk symbol next to it) due to its higher energy and identical coverage to the SHB phase. Putting the phases into context with the introduced growth model, the metastable SBW* phase is our target structure (T in Figure 1). The geometrically similar LBW phase acts as the connector polymorph (C), which becomes thermodynamically accessible at higher temperatures, while the SHB phase (S) is the thermodynamically stable phase under ambient conditions.

By leveraging these system properties, we demonstrate that the metastable SBW* structure can be kinetically accessed in kinetic Monte Carlo (kMC) growth simulations at conditions that are experimentally feasible. In total we conduct 100 kMC growth simulations on a periodically repeated 96 Å x 96 Å area realized by a 24x24 square grid with a lattice side length of 4 Å. We start the simulations on an initially empty surface at a temperature of 350 K and a molecular partial pressure of $10^{-6}$ bar. The slightly elevated temperature facilitates the formation of the thermodynamically stable phase, while still being at the low end of typical hot-wall epitaxy deposition procedures.[13] Under these conditions the standing herringbone (SHB) phase is thermodynamically stable and emerges on the surface. As expected from Ostwald's rule, the low coverage phase of lying molecules in the LHB pattern builds up first within one second, as shown in Figure 4a. Subsequently the transition to the thermodynamic stable SHB structure occurs, which is completed within several minutes. In principle this step is not needed for the formation of the metastable structure. We could have directly grown the LBW connector phase by starting at a high temperature where this structure is stable, but we want to demonstrate that without any changes in the deposition conditions the metastable SBW* structure is not accessible, and the system always ends up in the thermodynamic minimum.

Five Minutes ($t = 300\ s$) after the deposition starts, we change the chemical potential of the gas phase reservoir to a region where the LBW connector phase becomes stable. This is achieved by a linear temperature increase up to 520 K and a logarithmic pressure drop to $10^{-8}$ bar, similar to a post-deposition thermal treatment.[6] To keep the simulations realistic, we assume that the temperature change from 350 K to 520 K and the pressure drop from $10^{-6}$ to $10^{-8}$ bar is not instantaneous, but commences within a duration of 100 s, as shown in Figure 4b. At $t = 400\ s$ the system undergoes a quite sharp phase transition to the stable LBW connector phase, covering close to 100 % of the surface in this structure. As the last step, at $t = 500\ s$, the chemical potential is changed to a region where the standing structure is favored again, which is achieved by cooling the system down to ambient temperature (300 K) and increasing the partial pressure of the adsorbate to $10^{-6}$ bar. In contrast to the deposition phase during the beginning of the simulation, the system now transforms into the metastable SBW* polymorph. At $t = 3000\ s$, on average around 99 % of the surface is covered with our target phase (see red line in Figure 4a).

Having created the metastable phase, a pressing question for potential technological applications is how long it remains stable. Since the SBW* phase consists of only upright standing molecules only a desorption processe of a standing molecule can trigger the degradation of this structure. Exemplarily, at a temperature of 300 K the expected elapsed time for the next desorption process to happen is on average $25 \cdot 10^6$ years. In practice this can be seen as a stable structure.

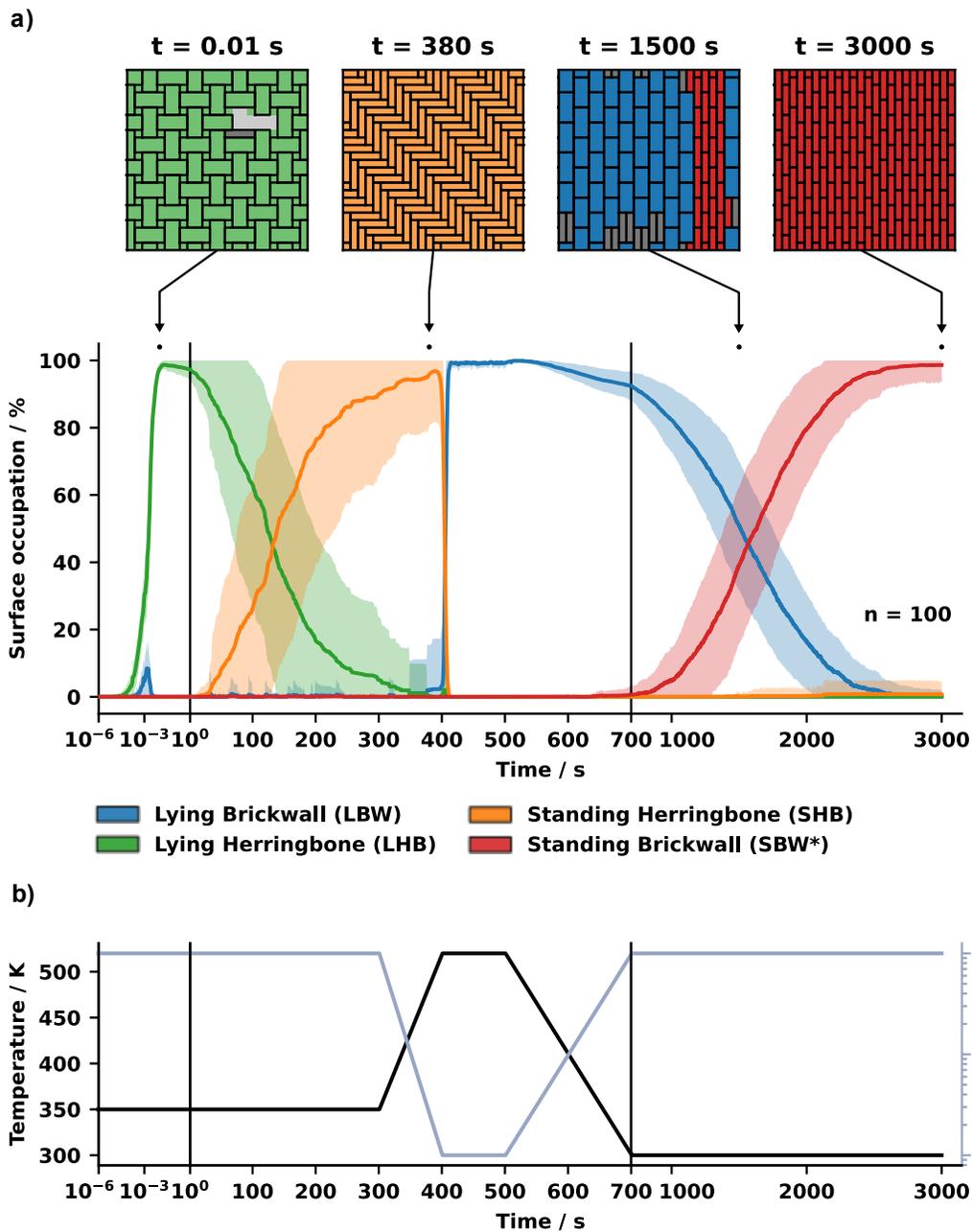

**Figure 4.** a) Surface occupation of the different surface structures averaged over 100 kMC runs on a 24 x 24 lattice. The shaded area of the curves represents the standard deviation calculated at each time point. The first second of the simulation is plotted in logarithmic scale to display the fast growth of the LHB phase. After that, a linear time axis is used with a different scale after $t = 700\ s$. Exemplary, the surface configurations of one of the 100 runs at four

representative points in time is plotted on top. b) The temperature and pressure protocol during the simulation.

In conclusion, we proposed a growth mechanism that allows the targeted formation of metastable surface structures for inorganic/organic interfaces accommodating adsorption geometries in lying and standing molecules, that exhibit a special connector structure with two main properties: First, its Gibbs free energy curve needs to intersect with the Gibbs free energy curve of the metastable target structure and, secondly, the transition process from the connector phase to this structure needs to be preferred compared to the transition back to the thermodynamic stable structure. For a proof of concept, we conducted kinetic Monte Carlo growth simulations of a simplified interface model of rectangular molecules. The results showed that after applying a specific temperature and pressure protocol to the system a significant area of the simulation cell is occupied with the anticipated metastable structure.

**Computational Method**

In the simulations, growth is limited to the monolayer regime, as we are mainly interested in designing wetting layers. In our computational experiments the structural evolution of surface adsorbates is modelled in the framework of the *kinetic Monte Carlo (kMC)* method,[14,15] as implemented in the code *kmos3*.[18,19] In comparison to *Molecular Dynamics*, *kMC* vastly coarse grains the complex dynamics on atomic scales to a stochastic model that draws and executes so-called *elementary processes*, such as adsorption, desorption and on-surface diffusion/reactions, with a probability proportional to their rate constants. After each step the simulation time is propagated by a Poisson distributed time step. In *kmos3* this is implemented using the *Direct method* (*Variable Step Size method*)[14,15], i.e. that calculates the time step $\Delta t$ based on the sum of all rate constants $k_{tot}$ of processes that are possible at the current step and a uniformly distributed random number $r \sim \mathcal{U}(0,1)$:

$$\Delta t = -\frac{\ln(1-r)}{k_{tot}} \tag{2}$$

However, for time-dependent rates as it is the case in this study, Equation (2) is only a good approximation if the generated time steps $\Delta t$ are much smaller than the variation time of the rate constants. This is not the case in our model. For a fully closed mono-layer with only upright standing molecules on the surface, only slow desorption processes are available leading to a small total rate $k_{tot}$ and consequently to large time steps $\Delta t$. To consider the time-dependence of the rate constants correctly the time steps $\Delta t$ must be drawn from an inhomogenous Poisson distribution.[20] The time steps can be determined by solving

$$\ln(1-r) + \int_0^{\Delta t} k_{tot}(t + \Delta t')d\Delta t' = 0 \tag{3}$$

for $\Delta t$.[15] We implemented an algorithm that solves Equation (3) at every kMC step in the used kMC framework *kmos3*.[18] A detailed description of it can be found in the Supporting Information.

The intermolecular interactions during the growth simulations are treated by the on-the-fly backend of *kmos3*, which evaluates the rate constants during the run. The common time disparity problem in kMC simulations is circumvented by applying the time acceleration algorithm of Dybeck et al.[21,22]

The adsorption rate of molecules on the kMC lattice is derived from classical collision theory of an ideal gas[11]. The impingement rate of particles on a surface unit cell with area $A_{uc}$ at a temperature $T$ and pressure $p$ is given by

$$k_{tot}^{ads} = \frac{pA_{uc}}{\sqrt{2\pi m k_B T}} \qquad (4)$$

where $m$ is the mass of the molecule (we used the mass of an 9-10 anthraquinone molecule $m = 3.458 \cdot 10^{-25}$ kg) and $k_B$ is the Boltzmann constant. This expression only describes the total number of particles per unit time that appear in the unit cell and does not tell anything about which adsorption geometry the molecule assumes after adsorbing. To account for that, it is necessary to estimate the probability $s_i$ that the impinging molecule adsorbs in geometry $i$. For simplicity, we assume that all $N$ different adsorption geometries that can occur in a surface unit cell have the same probability $s = 1/N$. Therefore, the adsorption rate $k_i^{ads}$ for one molecule ending up in geometry $i$ is given by

$$k_i^{ads} = s k_{tot}^{ads} = s \frac{pA_{uc}}{\sqrt{2\pi m k_B T}} \qquad (5)$$

The desorption rate constant $k_{des}$ can be determined by requiring to be consistent with thermodynamics. This requires the adsorption/desorption reaction channel to fulfill detailed balance leading to the expression[11]

$$k_i^{des} \approx k_i^{ads} \exp\left(\frac{E_i^{ads} + E^{int} - \mu(T,p)}{k_B T}\right) \qquad (6)$$

where $E_i^{\text{ads}} < 0$ is the adsorption energy of an isolated molecule in geometry $i$ on the surface, $E^{\text{int}}$ is the interaction energy with the neighboring molecules and $\mu(T,p)$ is the chemical potential of the molecular gas reservoir. This expression is only an approximation as we estimate the Gibbs free energy of the adsorbed molecule only with its adsorption energy and neglect the vibrational, work and entropic contributions as they are usually small (see Supporting Information).[11]

For the on-surface processes such as diffusion, rotation and reorientation processes we used an Arrhenius type rate equation. The rate constant $k_{ij}$ from adsorption geometry $i$ to $j$ is then given by

$$k_{ij} = f_{ij}\exp\left(-\frac{\Delta E_{ij} + \Delta\Delta E^{\text{int}}}{k_\text{B}T}\right) \tag{7}$$

where $f_{ij}$ is the attempt frequency, $\Delta E_{ij}$ the corresponding energy barrier and $\Delta\Delta E^{int}$ the correction of the energy barrier resulting from the interaction of neighboring molecules. This correction of the energy barrier is calculated according to the Bronsted-Evans-Polanyi principle[23–28] and reads

$$\Delta\Delta E^{\text{int}} = \alpha\left(E_{\text{fin}}^{\text{int}} - E_{\text{ini}}^{\text{int}}\right) \tag{8}$$

where $\alpha \in [0,1]$ is a process specific parameter, $E_{\text{ini/fin}}^{\text{int}}$ is the interaction energy of the initial/final state depending on the specific molecular neighborhood on the surface at the time when the process is executed. A detailed derivation can be found in the Supporting Information.

ASSOCIATED CONTENT

**Data Availability Statement:** All the data used in this work and the adapted kMC time increment algorithm code for time-varying process rates are available from the authors upon reasonable request.

**Supporting Information:** Discussion of the Gibbs free energy calculation of monolayers; derivation of the Bronsted-Evans-Polanyi principle of the investigated system; additional computational details of the kMC simulations and the surface composition evaluation;detailed description of the developed algorithm for time-dependent process rates in kMC (PDF)

AUTHOR INFORMATION

**Notes**

The authors declare no competing financial interests.

ACKNOWLEDGMENT

Funding through the projects of the Austrian Science Fund (FWF): MAP-DESIGN 10.55776/Y1157 and HiTeq 10.55776/I5170 is gratefully acknowledged. Computational results have been achieved in part using the Vienna Scientific Cluster (VSC). We gratefully acknowledge the support of Martin Deiml and Sebastian Matera concerning the *kmos3* simulation framework. We acknowledge fruitful discussions with R. K. Berger, B. Ramsauer and C. Wachter.

# Supporting Information for
# "Metastable Monolayer Formation through a Connector Structure"

Simon B. Hollweger, Anna Werkovits, Oliver T. Hofmann

Institute of Solid State Physics, TU Graz, Petersgasse 16/II, 8010 Graz, Austria

## S1. Gibbs free energy and chemical potential calculation

The surface Gibbs free energy per area $\gamma$ of an structure is given by [1]

$$\gamma = \frac{1}{A}(\Delta G(T) - \mu_{\text{mol}}(T,p)N) \tag{S1}$$

where $\Delta G$ is the Gibbs free energy difference between the clean surface and the surface with adsorbed molecules on it, $\mu_{\text{mol}}$ the chemical potential of one molecule in the gas reservoir at given temperature $T$ and partial molecular gas pressure $p$ and $N$ is the number of molecules adsorbed in the surface area $A$. For simplicity we approximated the Gibbs free energy $\Delta G$ with the adsorption energy $E^{\text{ads}}$ and ignore the entropic, work and vibrational contributions. The entropic and work contributions are usually small and negligible.[1] The vibrational contribution cancels out to a certain degree with the vibrational contribution of the chemical potential, and it is therefor justified to be neglected in both terms $\Delta G$ and $\mu_{\text{mol}}$. With all these approximations we end up with the equation introduced in the main text reading

$$\gamma = \frac{1}{A}(E^{\text{ads}} - \mu(T,p)N) \tag{S2}$$

For calculating the chemical potential $\mu$ an 9-10 anthraquinone molecule was used as reference (Figure S1). The chemical potential $\mu_{mol}$ was calculated according to the thermodynamical ideal gas laws, as it is common in ab-initio thermodynamics, for details compare Rogal and Reuter.[1] The chemical potential consists of several contributions[2], that read

$$\mu_{mol} = \mu^{trans} + \mu^{rot} + \mu^{elec} + \mu^{vib} + \mu^{nucl}. \tag{S3}$$

As discussed in the work of Rogal and Reuter[1] the contribution $\mu^{nucl}$ from the nuclei is usually small and can be neglected. The vibrational part $\mu^{vib}$ is also neglected for reasons already mentioned above. The electronic contribution is given by

$$\mu^{elec} = E_{mol}^{tot} - k_B T \ln(I^{spin}) \tag{S4}$$

where the total electron energy $E_{mol}^{tot}$ is set to zero as it just introduces an offset. The contribution from the spin degeneracy $I^{spin}$ is zero for our reference molecule. With this, the used approximated chemical potential $\mu$ is then given by[2,3]

$$\mu(T,p) = \mu^{trans} + \mu^{rot} \tag{S5}$$

With

$$\mu^{trans} = -k_B T \ln\left[\left(\frac{2\pi m}{h^2}\right)^{\frac{3}{2}} \frac{(k_B T)^{\frac{5}{2}}}{p}\right] \tag{S6}$$

and

$$\mu^{\text{rot}} = -k_B T \ln\left[\frac{\sqrt{\pi I_1 I_2 I_3}}{\sigma}\left(\frac{8\pi^2 k_B T}{h^2}\right)^{\frac{3}{2}}\right]. \tag{S7}$$

All the used molecular constants of Anthraquinone are listed in Table S1.

**Table S1.** Used parameters for chemical potential calculation

| Parameter | Symbol | Value |
|---|---|---|
| Mass | $m$ | $3.458 \cdot 10^{-25}$ kg |
| Moments of inertia | $I_1$ | $7.636 \cdot 10^{-45}$ kg m² |
| | $I_2$ | $1.895 \cdot 10^{-44}$ kg m² |
| | $I_3$ | $2.659 \cdot 10^{-44}$ kg m² |
| Symmetry number | $\sigma$ | 4 |

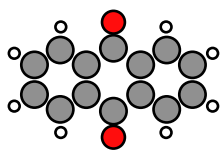

**Figure S1.** The used 9-10 anthraquinone reference molecule for calculating the chemical potential. Grey are carbon, white are hydrogen and red are oxygen atoms.

## S2. Bronsted-Evans-Polanyi principle

In general, the Bronsted-Evans-Polanyi[4–9] (BEP) principle states that the activation energy barrier $\Delta E$ and the enthalpy difference $\Delta H$ between initial and final state of reactions of similar type are in a linear relation. This is described by a proportionality parameter $\alpha \in [0,1]$ and a reference energy $E_\ddagger^R$.

$$\Delta E = \alpha \Delta H + E_\ddagger^R \tag{S8}$$

The parameters $\alpha$ and $E_\ddagger^R$ are usually determined by performing a linear fit on calculated transition energy barriers of different processes of the same reaction family (e.g. only diffusion processes).

For the kinetic Monte Carlo simulation, we need to know how the energy barrier $\Delta E$ of a process changes between the case where interacting neighbor molecules are present and the non-interacting single molecule case. We can use the Bronsted-Evans-Polanyi relation in Equation (S8) to write for the interacting transition barrier $\Delta E'$ and the non-interacting case $\Delta E^0$

$$\Delta E^0 = \alpha \underbrace{\left(E_{\text{fin}}^0 - E_{\text{ini}}^0\right)}_{\Delta H^0} + E_\ddagger^R \tag{S9}$$

$$\Delta E' = \alpha \underbrace{\left(E'_{\text{fin}} - E'_{\text{ini}}\right)}_{\Delta H'} + E_\ddagger^R \tag{S10}$$

where $E_{\text{ini}}$ and $E_{\text{fin}}$ are the initial and final energies of the molecule, see Figure S2. The change $\Delta \Delta E^{\text{int}}$ in the process energy barrier is now the difference between these two expressions and reads

$$\Delta\Delta E^{\text{int}} = \alpha[(E'_{\text{fin}} - E'_{\text{ini}}) - (E^0_{\text{fin}} - E^0_{\text{ini}})] =$$

$$= \alpha\left[\underbrace{(E^0_{\text{fin}} + E^{\text{int}}_{\text{fin}})}_{E'_{\text{fin}}} - \underbrace{(E^0_{\text{ini}} + E^{\text{int}}_{\text{ini}})}_{E'_{\text{ini}}} - (E^0_{\text{fin}} - E^0_{\text{ini}})\right] = \quad (S11)$$

$$= \alpha[E^{\text{int}}_{\text{fin}} - E^{\text{int}}_{\text{ini}}].$$

We found now the expression from the main text (Equation (8)) for correcting the non-interacting energy barrier $\Delta E^0$ to find the interacting energy barrier

$$\Delta E' = \Delta E^0 + \Delta\Delta E^{\text{int}}. \quad (S12)$$

We can see that only the interaction energies $E^{\text{int}}_{\text{fin}}$ and $E^{\text{int}}_{\text{ini}}$ of the molecules in the initial and final state are necessary to account for the effect of the interactions on the transition barrier.

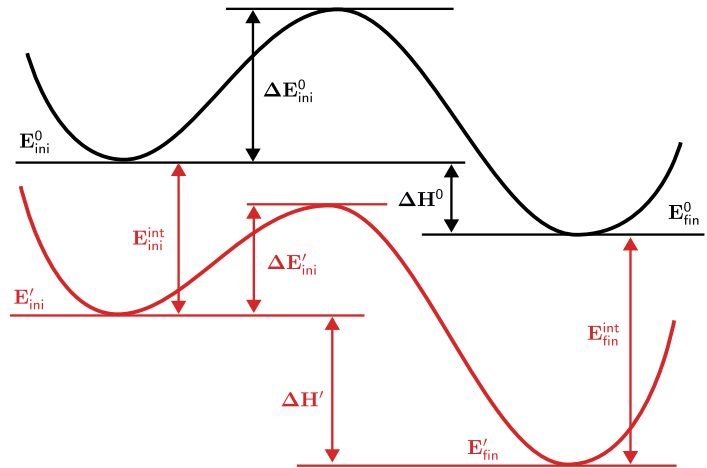

**Figure S2.** Reaction path of a single molecule on the surface with no interactions (black line) and the same process but with attractive neighbor interactions at initial and final state (red line).

These corrections are used in all the performed kinetic Monte Carlo simulations. For the reorientation processes from a face-on lying geometry to a long-edge standing orientation a scaling value of $\alpha = 0.7$ was used, for all other processes $\alpha$ was set to 0.5 .

### S3. Time varying rates in kinetic Monte Carlo

The simulation time in a kinetic Monte Carlo run is propagated according to a zeroth order Poisson process.[10,11] This means we draw the random time step $\Delta t$ from an exponential probability distribution given by

$$\Delta t \sim p(\Delta t | k_{\text{tot}}) = k_{\text{tot}} e^{-k_{\text{tot}} \Delta t} \tag{S13}$$

where $k_{\text{tot}}$ is the total sum of all process rates in the kMC system at the current kMC step. We can sample this distribution with a uniformly distributed random number $r \sim \mathcal{U}(0,1)$ by using

$$\Delta t = -\frac{\ln(1-r)}{k_{\text{tot}}} \tag{S14}$$

This probability distribution is only valid for constant rates $k_{\text{tot}}$. However, in this study we have to consider time varying rates in the system and we need to determine the time step $\Delta t$ from a distribution which considers the time dependence. The probability distribution for the general time dependent case at time point $t$ is given by [11,12]

$$\Delta t \sim p(\Delta t | t, k_{\text{tot}}) = k_{\text{tot}}(t + \Delta t) e^{-\int_0^{\Delta t} k_{\text{tot}}(t + \Delta t') d\Delta t'}. \tag{S15}$$

To sample this time dependent distribution with an uniformly distributed random variable $r \sim \mathcal{U}(0,1)$ we first need to calculate the cumulative distribution function

$$F_c(\Delta t|t, k_{\text{tot}}) = \int_0^{\Delta t} p(\Delta t'|t, k_{\text{tot}})d\Delta t' =$$

$$= \int_0^{\Delta t} k_{\text{tot}}(t+\Delta t')e^{-\int_0^{\Delta t'} k_{\text{tot}}(t+\Delta t'')d\Delta t''}d\Delta t'$$

(S16)

With the substitution $\mu = \int_0^{\Delta t'} k_{\text{tot}}(t+\Delta t'')d\Delta t''$ and $d\mu = k_{\text{tot}}(t+\Delta t')d\Delta t'$ we find

$$F_c(\Delta t|t, k_{\text{tot}}) = \int_{\mu(0)}^{\mu(\Delta t)} e^{-\mu} d\mu =$$

$$= -e^{-\mu}\Big|_{\mu(0)=0}^{\mu(\Delta t)=\int_0^{\Delta t} k_{\text{tot}}(t+\Delta t')d\Delta t'} =$$

(S17)

$$= 1 - e^{-\int_0^{\Delta t} k_{\text{tot}}(t+\Delta t')d\Delta t'}.$$

With the inverse cumulative distribution function $F_c^{-1}$ and the uniformly distributed $r \sim \mathcal{U}(0,1)$ we can find now the random time step $\Delta t$ following the time dependent distribution function $p(\Delta t|t, k_{\text{tot}})$ in Equation (S15)

$$\Delta t = F_c^{-1}(r).$$

(S18)

Unfortunately, it is not possible to find a closed form for $F_c^{-1}$. We can only write an implicit equation of the form

$$\ln(1-r) + \int_0^{\Delta t} k_{\text{tot}}(t+\Delta t')d\Delta t' = 0$$

(S19)

that needs to be solved for $\Delta t$. Notice if $k_{\text{tot}}(t)$ is changing only slowly in the time interval $(t, t+\Delta t)$ we can approximate it with a constant $k_{\text{tot}}$ and can use the equation for the time-independent case mentioned in Equation (S14). However, as already described in the main text, there exist cases where the time step $\Delta t$ is large and this approximation is not justified. To account for this, we adapted the time propagation algorithm in *kmos3* and solved the implicit Equation (S19) on a time grid $\tau_i$ that is fine enough to represent the time dependent total rate $k_{\text{tot}}$ sufficiently exact with a step function. In case of a step function for $k_{\text{tot}}$ the integral in Equation (S19) can be rewritten as a sum over rectangles and can be solved iteratively. If we assume that the current simulation time $t_{\text{kMC}}$ lies within the time grid interval $(\tau_{i-1}, \tau_i)$, we can write for the integral in Equation (S19)

$$\int_0^{\Delta t} k_{tot}(t_{kMC} + \Delta t')d\Delta t' \approx$$

$$\approx \begin{cases} k_{\text{tot}}(\tau_{i-1})\Delta t & \text{for } t_{\text{kMC}} + \Delta t \leq \tau_i \\ k_{\text{tot}}(\tau_{i-1})[\tau_i - t_{\text{kMC}}] + & \text{for } \tau_i < t_{\text{kMC}} + \Delta t \leq \tau_{i+1} \\ \quad + k_{\text{tot}}(\tau_i)[t_{\text{kMC}} + \Delta t - \tau_i] & \\ \vdots & \vdots \\ k_{\text{tot}}(\tau_{i-1})[\tau_i - t_{\text{kMC}}] + & \text{for } \tau_{i+N} < t_{\text{kMC}} + \Delta t \leq \tau_{i+N+1} \\ \quad + [\sum_{k=1}^{N} k_{\text{tot}}(\tau_{i+k-1})[\tau_{i+k} - \tau_{i+k-1}]] & \\ \quad + k_{\text{tot}}(\tau_{i+N})[t_{\text{kMC}} + \Delta t - \tau_{i+N}] & \end{cases} \quad (S20)$$

With this approximation at hand, we can formulate an algorithm that gives us the time step $\Delta t$ that solves Equation (S19). We simply have to insert the individual cases from Equation (S20) into (S19) starting from the top and solve for $\Delta t$. We can stop this iterative scheme as soon as a found time step $\Delta t$ fulfills the case condition or the final simulation time $t_{\text{final}}$ is exceeded. In practice the following steps are required:

1. Compute the time step corresponding to the first case in (S20) that gives

$$\Delta t = -\frac{\ln(1-r)}{k_{\text{tot}}(\tau_{i-1})} \qquad (S21)$$

2. Then check if $t_{\text{kMC}} + \Delta t \leq \tau_i$, if yes, we stop and can propagate the simulation time with $\Delta t$. This is just a regular kMC time step as it would be the case for the time independent case.

3. If the determined $\Delta t$ in step 1 overshoots the current time interval we initialize a counter $N = 0$ and propose a new time step with

$$\Delta t = \tau_{i+N} - t_{\text{kMC}} + \frac{-\ln(1-r) - k_{\text{tot}}(\tau_{i-1})[\tau_i - t_{\text{kMC}}] - \sum_{k=1}^{N} k_{\text{tot}}(\tau_{i+k-1})[\tau_{i+k} - \tau_{i+k-1}]}{k_{\text{tot}}(\tau_{i+N})} \qquad (S22)$$

4. Now we need to check again if $t_{\text{kMC}} + \Delta t \leq \tau_{i+N+1}$. If this is fulfilled, we are done and found a proper time step. If it is not fulfilled, we increment the counter ($N = N + 1$) and repeat this step until it is fulfilled. Notice that in the first iteration with $N = 0$ the sum term in Equation (S22) is empty and therefore zero. We also terminate the time step search if we exceed the final simulation time $t_{\text{final}}$.

In Figure S3 a flowchart of this algorithm for determining $\Delta t$ can be found.

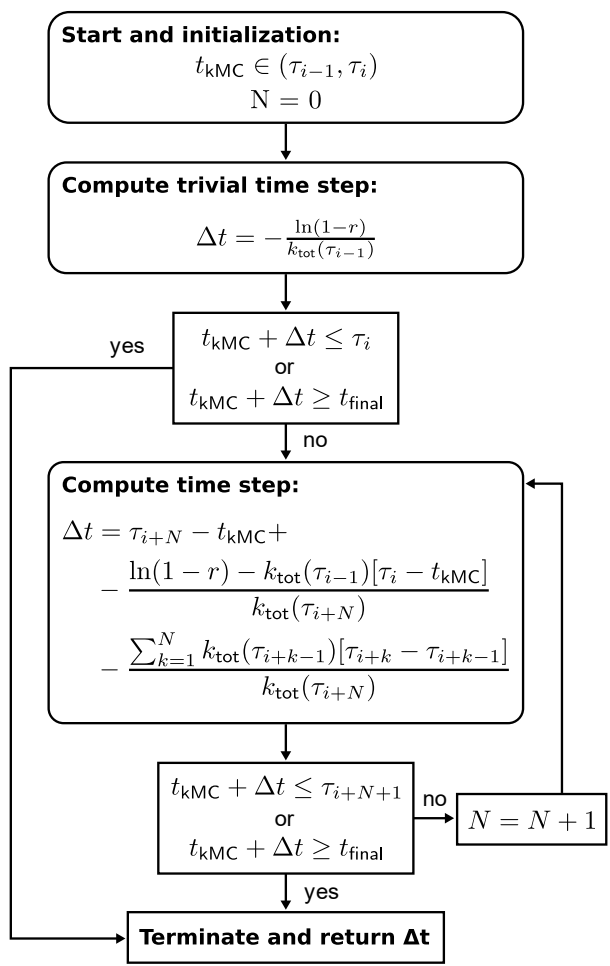

**Figure S3.** Algorithm for determining the kMC time step in the case of time varying rate constants

With that algorithm we can propagate the simulation time in our model even for large time jumps that are in the range of the variation times of the rate constants.

## S4. Kinetic Monte Carlo simulation setup

The kinetic Monte Carlo simulations were performed with an adapted version of the *kmos3* simulation package.[13] The kMC models were generated with the on-the-fly (otf) backend that considers the intermolecular interactions during runtime. For the interactions only the nearest neighbors were considered, as it can be seen in Figure 2b in the main text. Additionally, the previously described algorithm for determining the time step in the case of time dependent rate constants was implemented in *kmos3*[14] by us and used in all simulation runs. All the parameters used for the time acceleration scheme[15] implemented in *kmos3* are listed in Table S2.

**Table S2.** Used parameters for the time acceleration scheme. The names are the ones used in the *kmos3* code documentation[14].

| Parameter | Value |
|---|---|
| buffer_parameter | 500 |
| sampling_steps | 20 |
| execution_steps | 200 |
| threshold_parameter | 0.2 |

All possible elementary processes except the adsorption and desorption processes are visualized in Figure S4. The used energy barriers needed for evaluating the Arrhenius equation are listed in table Table S3. All attempt frequencies for all elementary processes were set to $f = 10^{12}\ s^{-1}$.

**Table S3.** All barriers used for calculating the on-surface processes visualized in Figure S4. The corresponding attempt frequencies are all equal and set to $f = 10^{12}\ s^{-1}$.

| Parameter | Value |
|---|---|
| Lying diffusion barrier $\Delta E_L^{diff}$ | 0.3 eV |
| Standing diffusion barrier $\Delta E_S^{diff}$ | 0.3 eV |
| Stand up barrier $\Delta E_{LS}^{reor}$ | 0.45 eV |
| Fall-over barrier $\Delta E_{SL}^{reor}$ | 0.25 eV |
| Lying rotation barrier $\Delta E_L^{rot}$ | 0.3 eV |
| Standing rotation barrier $\Delta E_S^{rot}$ | 0.3 eV |

**Face-on lying diffusion**

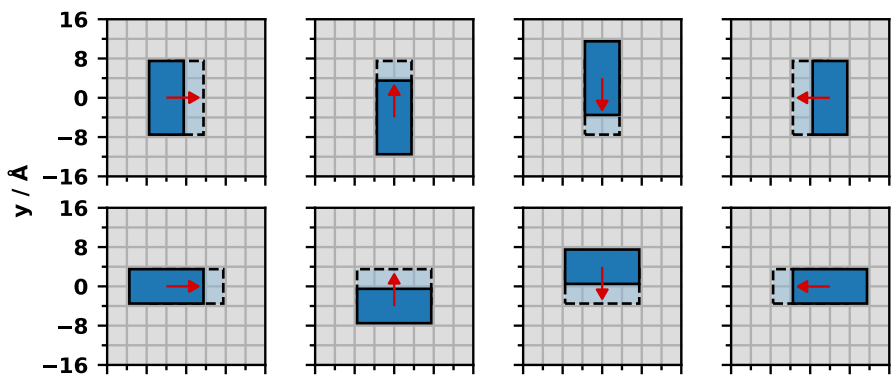

**Lying-Standing reorientation**

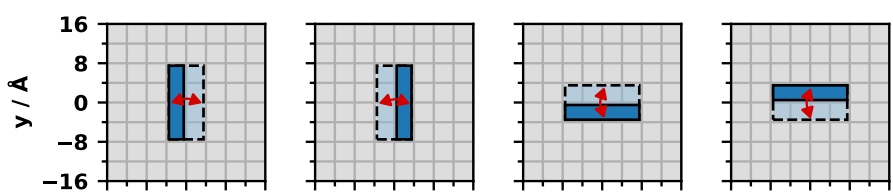

**Face-on lying rotation**

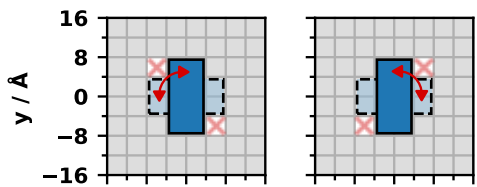

**Upright standing diffusion**

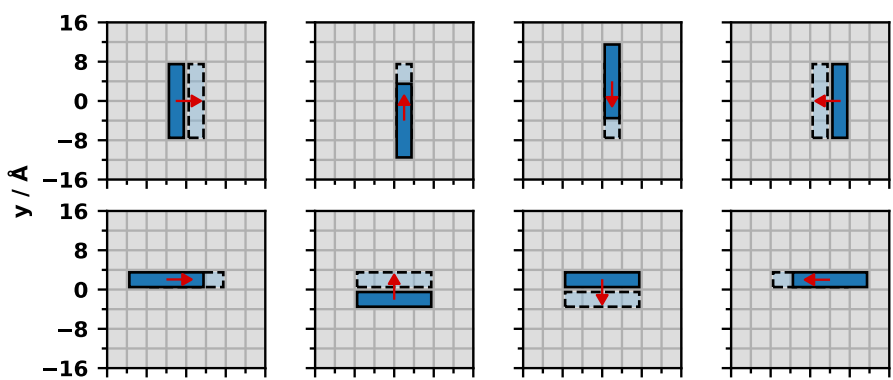

**Upright standing rotation**

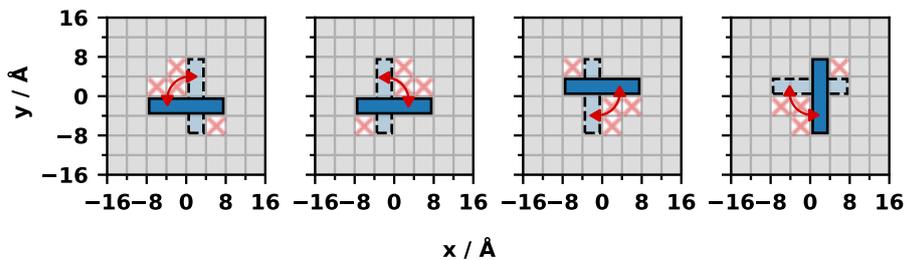

**Figure S4.** All possible elementary processes and the corresponding requirements for empty sites for the process to be possible.

## S5. Surface composition evaluation

For evaluating the areal occupation of the different polymorphs on the surface during a kMC run a simple expansion algorithm was used.

1. The first step is to iterate through the surface sites and check if one can find a molecular neighborhood that is identical to a unit cell of the structures one is checking for.

2. If a unit cell was found, we try to expand it by checking for neighbor molecules that obey the periodicity of the found unit cell. If a neighbor molecule fits in the structure, the search is continued from this molecule onwards. The expansion of this domain stops if no neighbors could be found anymore.

3. As soon as a domain expansion is completed one checks if it satisfies a minimum size requirement to be considered as domain (e.g. number of domain molecules $n > n_{\min}$). Then we go back to step 1 and further iterate the lattice sites to search for the next unit cell to be expanded. All molecules that are already categorized are skipped.

When every lattice site and unit cell type is checked we end up with a list of domains that can be summed up to give the areal occupations of the different structures. The minimum size to be considered as a domain of a structure, $n_{\min} = 4$ was used for lying structures and $n_{\min} = 6$ for standing structures.

In the histogram in Figure S5 the final yield (at $t = 3000$ s) of the target SBW* structure of all 100 performed kMC simulations is shown. The average yield is around 99 %.

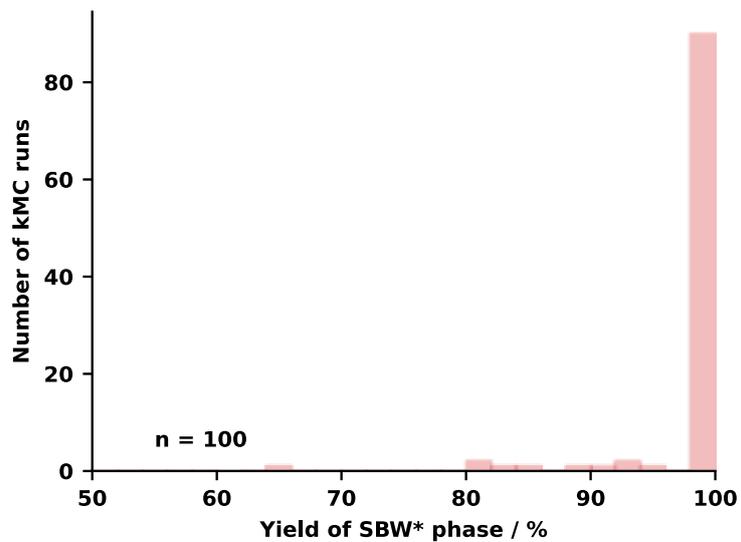

**Figure S5.** Histogram of the number of kMC simulations binned by the yield of the metastable SBW* phase after $t = 3000$ s.

In Figure S6 the time evolution of the mean Gibbs free energy of adsorption of the system is visualized.

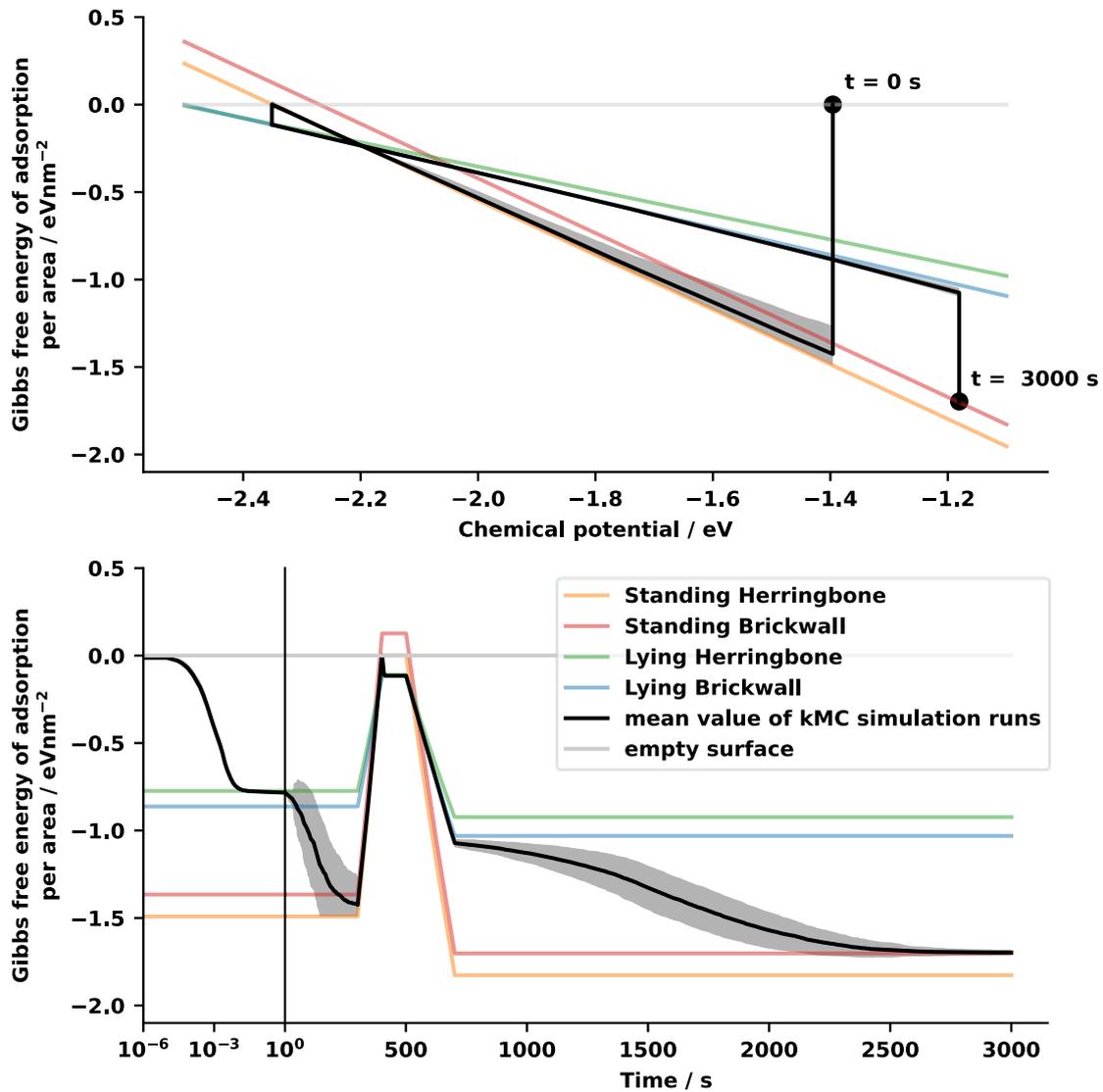

**Figure S6.** Time evolution of the mean Gibbs free energy of adsorption per area of all 100 runs: (a) the energy trajectory plotted over the chemical potential of the molecular gas reservoir; (b) the energy trajectory plotted over the simulation time

## S6. Lifetime estimation

To estimate the stability of the SBW* phase it is sufficient to focus on the rate constant of a standing molecule within the SBW* lattice desorbing from the surface. This type of process is the only available process in the situation of a fully closed standing monolayer and therefore the only one capable of triggering a phase transition to the thermodynamic stable SHB phase. The average waiting time $\langle t_{\text{des}} \rangle$ until the next desorption process is happening can be estimated by determining the expectation value of the underlying Poisson distribution of this process and is given by $\langle t_{\text{des}} \rangle = 1/k_{\text{des}}$. The desorption rate $k_{\text{des}}$ is given by Equation (6) of the main text. At a temperature of $T = 300$ K (pressure $p$ is not relevant as it cancels between $k_{\text{ads}}$ and $\mu^{\text{trans}}$ in Equation (6)) we find

$$\langle t_{\text{des}} \rangle = 7.93 \times 10^{14} \text{ s} = 25.15 \times 10^6 \text{ years}. \tag{S23}$$

This sufficiently shows the relative stability of the metastable SBW* phase.